# Unveiling pseudospin and angular momentum in photonic graphene

Daohong Song[1], Vassilis Paltoglou[2], Sheng Liu[3,4], Yi Zhu[5], Daniel Gallardo[3], Liqin Tang[1], Jingjun Xu[1#], Mark Ablowitz[6], Nikolaos K. Efremidis[2], and Zhigang Chen[1,3*]

[1]The MOE Key Laboratory of Weak-Light Nonlinear Photonics, and TEDA
Applied Physics Institute and School of Physics, Nankai University, Tianjin 300457, China
[2]Department of Applied Mathematics, University of Crete, 71409 Heraklion, Crete, Greece
[3]Department of Physics and Astronomy, San Francisco State University, San Francisco, CA 94132
[4]The MOE Key Laboratory of Space Applied Physics and Chemistry,
and Shaanxi Key Laboratory of Optical Information Technology, School of Science,
Northwestern Polytechnical University, Xi'an 710129, China
[5]Zhou Pei-Yuan Center for Applied Mathematics, Tsinghua University, Beijing, 100084, China
[6]Department of Applied Mathematics, University of Colorado, 526 UCB, Boulder, CO 80309

*zhigang@sfsu.edu,  #jjxu@nankai.edu.cn

**Pseudospin, an additional degree of freedom inherent in graphene, plays a key role in understanding many fundamental phenomena such as the anomalous quantum Hall effect, electron chirality and Klein paradox. Unlike the electron spin, the pseudospin was traditionally considered as an unmeasurable quantity, immune to Stern-Gerlach-type experiments. Recently, however, it has been suggested that graphene pseudospin is a real angular momentum that might manifest itself as an observable quantity, but so far direct tests of such a momentum remained unfruitful. Here, by selective excitation of two sublattices of an artificial photonic graphene, we demonstrate pseudospin-mediated vortex generation and topological charge flipping in otherwise uniform optical beams with Bloch momentum traversing through the Dirac points. Corroborated by numerical solutions of the linear massless Dirac-Weyl equation, we show that pseudospin can turn into orbital angular momentum completely, thus upholding the belief that pseudospin is not merely for theoretical elegance but rather physically measurable.**

Graphene, a two-dimensional (2D) honeycomb lattice of carbon atoms, has been highly touted and tested as an extraordinary material for many applications, apart from elucidating fundamental phenomena in quantum and condensed matter physics [1-4]. Central to this enthusiasm is the unique electronic band structure of the graphene lattice, exhibiting a linear energy dispersion relation in the vicinity of the so-called Dirac points (Fig. 1), where electrons behave as massless relativistic particles. While extracting a single atomic layer of graphite (as "natural" carbon-based graphene) can be readily accomplished nowadays with simple laboratory techniques[3], there is a surge of interest recently in creating "artificial" graphene systems; not only for electrons but also for atoms, photons, and polaritons, from nanopatterning of 2D electron gases to assembling molecules on metal surfaces, and from trapping ultracold atoms in optical lattices to engineering coupled micropillars in a semiconductor microcavity [5-13]. This is simply because artificial graphene can provide a tunable platform to explore physical phenomena that are otherwise difficult or impossible to achieve in natural graphene. In particular, photonic graphene (a honeycomb array of evanescently coupled waveguides[14]) has proven to be a useful tool for investigating graphene physics in various optical settings[15-23]. Since photonic lattices offer exquisite control over initial conditions and allow for monitoring the actual wavefunction (including phase), it is possible to directly observe graphene wave dynamics using classical light waves in regimes not accessible in natural graphene. Exemplary successes include the recent demonstrations of defect-free Tamm-like edge states[20,21], strain-induced pseudomagnatic fields and photonic Landau levels[22], and the photonic Floquet topological insulators[23].

In this Letter, by use of photonic graphene as a test bed, we investigate and unveil yet another intriguing phenomenon -- the "pseudospin." While the concept of pseudospin was introduced initially due to the mathematical analogy between the graphene sublattice degree of freedom and the electron spin in the original Dirac equation, it has certainly turned into one of the paradigms of graphene physics[1-4]. Although recent experiments with angle-resolved photoemission spectroscopy has led to direct probing of the Berry phase in graphene systems[24], the pseudospin itself is considered to be unmeasurable. Unlike the electron spin, even if the pseudospin corresponds to an angular momentum (AM), it is not detectable by any magnetic field[25,26]. In our optical setting, however, we can selectively excite each of the two sublattices of photonic graphene, breaking the degeneracy introduced by the two inequivalent atomic sites in the HCL, and thus uncovering the underlying physics of pseudospin. Specifically, we employ two different methods to alternatively excite one of the two sublattices forming the photonic graphene, and observe in both real and momentum spaces the vortex generation when an initially vortex-free probe beam with Floquet momentum in the vicinity of the Dirac points travels through the lattice. Moreover, the topological charge of the generated vortex flips as the excitation of the probe beam moves from one sublatttice to another. By comparing our experimental results with numerical and theoretical analyses of the linear massless Dirac-Weyl equation, we show that the observed vortices (or optical beams carrying orbital AM[27]) are a direct consequence of the AM transfer from the lattice to the probe beam. Unlike the electron spin, such pseudospin AM is not associated with any intrinsic property of particles, but rather arises from the substructure in space (sublattices) that the particles (or wave packets) live in. Our work may lead to new insights of pseudospin-mediated fundamental phenomena in both natural and artificial graphene systems.

The honeycomb lattice (HCL) is composed of two inter-penetrating triangular sublattices, whose representing lattice sites are denoted by A and B as shown in Fig. 1a. The band gap structure $\beta(k_x, k_y)$ plotted in Fig. 1b is calculated from the following paraxial Schrödinger-type equation describing light propagation in the photonic lattice[20]:

$$i\frac{\partial \Psi(x,y,z)}{\partial z} = -\frac{1}{2k_0}\nabla^2\Psi(x,y,z) - \frac{k_0 \Delta n(x,y)}{n_0}\Psi(x,y,z) \equiv H_0\Psi, \quad (1)$$

where $\Psi$ is the electric field envelope of the probe beam, $z$ is the longitudinal propagation distance, $k_0$ is the wavenumber, $n_0$ is the background refractive index of the medium, and $\Delta n$ is the induced index change forming the HCL. In Eq. (1), $H_0$ is the continuous Hamiltonian of the system, whose eigenvalues are the wavenumbers along the $z$-direction (i.e., the propagation constant $\beta$). From Fig. 1b, one can see clearly the touching of two bands at the six Dirac points, where the Floquet-Bloch dispersion relation is linear (Fig. 1c). These Dirac points are located at the corners of the first Brillioun zone (BZ) of the HCL, noted as $K$ and $K'$ in Fig. 1d. Applying the coupled-mode theory (under the tight-binding approximation) to Eq. (1), one can obtain a two-band simplified description of the paraxial model[17]. In the continuous limit and for excitations near the Dirac points, the coupled mode equation turns into

the linear Dirac equations typically used for describing massless Dirac particles in graphene:

$$i\partial_z \psi_A + (\partial_x - i\mu\partial_y)\psi_B = 0$$
$$i\partial_z \psi_B - (\partial_x + i\mu\partial_y)\psi_A = 0 \quad , \quad (2)$$

where $\mu = (-1)^m = \pm 1$, and $m = 0,\ldots,5$ is the index of the six Dirac points shown in Fig. 1d. The associated Hamiltonian can be written as $H = \sigma_y p_x - \mu\sigma_x p_y$, where $\boldsymbol{p} = (p_x, p_y)$ is the Bloch momentum measured from the Dirac points, $\boldsymbol{\sigma} = (\sigma_x, \sigma_y)$ are the Pauli matrices[3]. Detailed derivations and analysis of Eq. (2) are presented in the Supplementary Information section. Thus, an optical beam with Bloch momentum at the close vicinity of Dirac points is governed by the Dirac equation, akin to massless Dirac Fermions in graphene. The amplitude of the optical wave in the spatially separated sublattice sites (marked as A and B in Fig. 1a) is modeled by the two-component spinor function $\psi_A$ and $\psi_B$, respectively. Therefore, as we shall elaborate below, the sublattice states play the role of electron spins, typically referred to as "pseudospin". In light of such an analogy, a natural question arises: is the lattice spin associated with real AM[25] observable in our optical setting?

To answer the above question, we first perform a numerical beam propagation simulation of the paraxial equation [Eq. (1)] to illustrate the pseudospin-mediated vortex generation by sending three interfering plane waves as a probe to the HCL (Fig. 2a). The three input wave vectors point at three alternative Dirac points (*K* or *K'*), thus the two sublattices can be selectively excited by the probe beam which exhibits a triangular lattice pattern (Fig. 2e). Specifically, in the first setting (three waves of equal phase), only sublattice A is excited (Fig. 2b). In the second setting (three waves of a $2\pi/3$ phase difference in ***k***-space), only sublattice B is excited but not sublattice A (Fig. 2f). Surprisingly, although the output intensity of the probe beam displays a similar conical diffraction pattern[14,28] as shown in Figs. 2c and 2g, their phase structure as monitored from interferograms is dramatically different. In both cases, a global singly-charged vortex is created, as identified by a fork bifurcation in the central fringes, although the topological charges are opposite (Figs. 2d, 2h). It should be noted that the input beam initially contains no phase singularity. In addition, when both sublattices A and B are simultaneously excited or when the HCL is replaced by a single triangular lattice, no vortex is generated. These results suggest that the vortex generation and topological charge flipping is a direct consequence of the special symmetry and sublattice degree of freedom of the HCL, which will be confirmed later by introducing the total AM and directly solving the Dirac equation.

Next, we experimentally demonstrate the pseudospin-mediated vortex generation in a photonic graphene system – the HCL is created by optical induction which translates lattice intensity pattern into refractive index change in a photorefractive nonlinear crystal[29-32]. A detailed description of the experimental setup is given in the Supplementary Information section. Typical results are shown in Fig. 3 and Fig. 4, which correspond to two different methods of selectively exciting the two graphene sublattices. The index change associated with the HCL is about $1.5 \times 10^{-4}$, and the lattice constant is about 7μm (Fig. 3a). The BZ spectrum of the induced lattice shown in Fig. 3b is measured separately using BZ spectroscopy with incoherent light[33]. In the first method, three broad Gaussian beams forming a triangular lattice pattern (Fig. 3e) are carefully aimed onto the three Dirac points in the first BZ (Fig. 3f). To selectively excite the two sublattices in the same experimental setting, the probe lattice has the same period as the sublattices and it can be readily translated along the transverse direction by moving the focus lens. Note that the intensity/polarization of all beams is chosen such that there is no nonlinear self-action of the probe beam. Clearly, when the probe beam excites only sublattice A (Fig. 3c) or sublattice B (Fig. 3g), there is not much difference in the output intensity pattern after propagating 2 cm through the HCL, as in both cases the beam exhibits a low intensity in central region due to conical diffraction. However, the interferograms obtained with an inclined reference plane wave indicate that not only is a singly-charged optical vortex created in each case, but also the topological charge is flipped (opposite fringe bifurcation) when the excitation shifts from sublattice A to B (Figs. 3d, 3h). Under the same excitation condition, if the HCL is reconfigured into a single triangular lattice (i.e., only one sublattice is present), no fringe bifurcation is observed in the interferograms. Likewise, when the HCL is completely blocked, the three input beams propagate independently and no vortex is observed whatsoever. These simple tests indicate clearly that the observed vortex is not in any way related to experimental artifacts. In fact, our experimental observations agree well with the simulation results of Fig. 2.

In the second method, only two interfering beams are used as the probe, so only two Dirac points are initially excited (Fig. 4a). Nevertheless, due to the HCL symmetry and Bragg reflection, a new spectrum component emerges at the corresponding third Dirac point (Fig. 4e). Amazingly, the far-field intensity pattern of this new component (after a Fourier transform from momentum *k*-space back to real space) exhibits opposite vortex singularities as the two beams selectively excite sublattice A or B, revealed by the phase pattern obtained from both experimental interferogram (Figs. 4c, 4g) and numerical simulation (Figs. 4d, 4h). These results indicate again that the observed vortices and associated charge flipping arise from the honeycomb sublattice degree of freedom, i.e., the pseudospin.

To gain further insight of the underlying physics of pseudospin and unveil its angular momentum, we directly analyze the normalized Dirac equation [Eq. (2)]. Unlike the single-wavefunction description of the paraxial Schrödinger equation [Eq. (1)], the wave dynamics in the Dirac system is directly mapped into the two-component spinor wavefunctions $\psi_A$ and $\psi_B$. Consequently, there is absolutely no meaning in any form of interference between these two components. However, it is physically relevant to separate the wavefunction of the paraxial equation $\Psi(\mathbf{r},z)$ discretely according to its spatial location as $\Psi_A(\mathbf{r},z) = \Psi(\mathbf{R}^A_{m,n},z)$ and $\Psi_B(\mathbf{r},z) = \Psi(\mathbf{R}^B_{m,n},z)$, where $\mathbf{R}^{\{A,B\}}_{m,n}$ are the position vectors of the sublattice elements A and B with indices $(m,n)$ located in the same Wigner-Seitz cell as $\mathbf{r} = \hat{\mathbf{x}}x + \hat{\mathbf{y}}y$. We introduce the total angular momentum (AM) along *z*-direction as $J = L + S$, where $L = (\mathbf{r} \times \mathbf{p}) \cdot \hat{\mathbf{z}}$ is the orbital AM, $S = \mu \sigma_z / 2$ is the pseudospin ($\mu = 1$ or $-1$, and $\sigma_z$ is the respective Pauli matrix), $\hat{\mathbf{p}} = -i\nabla$. The average total lattice AM with respect to the spinor wavefunctions $\psi_A$ and $\psi_B$ is then given by

$$\langle J \rangle = \langle L \rangle + \langle S \rangle = \iint_{\mathbb{R}^2} \left[ -i\psi_A^* \frac{\partial}{\partial \phi} \psi_A - i\psi_B^* \frac{\partial}{\partial \phi} \psi_B + \frac{\mu}{2} \left( |\psi_A|^2 - |\psi_B|^2 \right) \right] dx\, dy \ . \qquad (3)$$

As an example, when $\mu=1$, any excitation of sublattice A (or B) will have a positive (or negative) contribution to the total AM of the system.

Typical numerical and analytical results obtained from the above Dirac equation are presented in Figs. 5a and 5b. Details of the calculations are given in the Supplementary Information section. To selectively excite a pseudospin eigenstate, we first separate the wavefunctions and the Floquet-Bloch modes discretely according to its spatial location in the two sublattices, and then solve the Dirac equation numerically with a Gaussian modulation of the separated Bloch modes as initial condition: $\Psi_{\{A,B\}}(\mathbf{r}, z=0; \mu) = u_{\mathbf{K}_\mu}^{\{A,B\}}(\mathbf{r}) e^{-(r/r_0)^2/2}$, $\Psi_{\{B,A\}}(\mathbf{r}, z=0; \mu) = 0$, where $u_{\mathbf{K}_\mu}^{\{A,B\}}$ are the Floquet-Bloch modes localized at $\{A,B\}$ lattice sites at $\mathbf{K}_\mu$. Note that close to the six Dirac points $m=1,\ldots,6$ the Floquet modes are degenerate when $m=2n$, and $m=2n+1$. Furthermore, even/odd values of $m$ characterize opposite local phase structures such that $u_{\mathbf{K}_{2m}}^{\{A,B\}} = (u_{\mathbf{K}_{2m+1}}^{\{A,B\}})^*$. The respective optical field in the Dirac limit is then given by $\psi_{\{A,B\}}(\mathbf{r}, z=0) = e^{-(r/r_0)^2/2}$ along with $\psi_{\{B,A\}}(\mathbf{r},z)=0$. At the input plane, the pseudospin is the only term that contributes to the total AM. However, from both the operator perspective (i.e. expressing $(\partial_x + i\mu\partial_y)$ in polar coordinates) and utilizing the Green's function, it is clear that when the sublattice A (or B) is initially excited, the field in sublattice B (or A) will be generated with a topological charge $\mu$ (or $-\mu$). This is clearly illustrated in the numerical results of Fig. 5a. The asymptotic structure of the beam is derived by a combination of the steepest descent and stationary phase approximations [as detailed in the supplementary material]. Specifically, when the sublattice A is initially excited, the analysis leads to

$$\psi_A(\mathbf{r}, Z) = F_+(r)e^{-(Z+r)^2/(4r_0^2)} + F_-(r)e^{-(Z-r)^2/(4r_0^2)} \tag{4}$$

$$\psi_B^{(\mu)}(\mathbf{r}, Z) = \mu e^{i\mu\phi}\left[G_+(r)e^{-(Z+r)^2/(4r_0^2)} + G_-(r)e^{-(Z-r)^2/(4r_0^2)}\right] \tag{5}$$

which confirms the vortex generation in sublattice B. Likewise, when sublattice B is initially excited, due to the transformation $\psi_A \leftrightarrow \psi_B, x \to -x, y \to y$ (holding for the Dirac system and thus for the above formulas), the optical field $\psi_A$ is generated with opposite vorticity $-\mu$. As shown in Fig. 5b, the asymptotic calculations are in excellent agreement with numerical results. Importantly, both our calculation and simulation show that, at propagation distances where the conical diffraction becomes appreciable, the amplitude profiles of the two spinor components become identical, thus resulting in zero lattice spin as also seen from Eq. (3). In fact, the initial pseudospin is completely transferred to the final orbital AM of the system, i.e., $\langle S \rangle_i = \langle L \rangle_f$. This explains the optical vortex (AM) generation in the HCL observed in our experiment. To substantiate our argument, we revisit the Schrödinger equation [Eq. (1)] with a HCL potential but perform numerical computations following the above procedure of decomposing the optical field into its spinor components. The results are shown in Figs. 5(c-h), where the interferograms of the two components are examined separately. Evidently, when the spinor state A (or B) is excited, the initial positive (or negative) value of the total pseudospin is converted to a vortex AM with a topological charge +1 (or -1) carried completely by sublattice B (or A).

Before closing, we emphasize that the total AM of the system $J$ would not be conserved should we ignore the pseudospin, simply because the orbital AM in the HCL is not conserved. In fact, the pseudospin represents the hidden AM due to the sublattice degree of freedom in the Dirac system of the HCL. The transverse derivatives associated with the Dirac equation [Eq. (2)] can be written as $\partial_x \pm i\partial_y = e^{\pm i\theta}(\partial_r - (i/r)\partial_\theta)$, thus if one spinor component does not carry vorticity, the second spinor component is going to be "compatible" only if its topological charge is +1 or -1. We note that our vortex generation and topological charge flipping is achieved by exciting the two sublattices at the same K valley (see Fig. 3) in a uniform (non-strained) HCL, thus no pseudomagnetic field is involved[22]. The observed pseudospin AM does not result from the valley-dependent nonzero Berry curvature at the K and K' valleys[34]. Experimentally, by exciting the same sublattice but from the two different sets of the valleys, we were unable to see any difference. Finally, we also mention that the concept of pseudospin could be extended to other types of lattices, such as the Kagome and Lieb lattices. In fact, with the Lieb lattices, it was recently suggested that the pseudospin is not merely a mathematical formality but rather has a physical effect[35].

In summary, we have demonstrated both theoretically and experimentally the pseudospin-mediated vortex generation in photonic graphene. Our results indicate clearly that the pseudospin is of real angular momentum, observable and measurable. Since this angular momentum arises as a direct outcome of the Dirac equation also widely studied in graphene systems and topological insulators, we envisage our results will have broader impact to other branches of physics and material sciences. In addition, our work also brings about a new mechanism to generate optical vortices which may find applications in photonics.

## Acknowledgements

The work is supported by the 973 programs (2013CB632703, 2013CB328702, 2012CB921900), the National Natural Science Foundation (11304165, 11204155, 61205001), PCSIRT（IRT0149） and 111 Project (No. B07013) in China, and by the Air Force Office of Scientific Research and the National Science Foundation in U.S. VP and NKE are supported by the action "ARISTEIA" in the context of the Operational Programme "Education and Lifelong Learning" that is co-funded by the European Social Fund and National Resources. We thank M.C. Rechtsman and M. Segev for discussion.

## Contributions

All authors contributed significantly to this work.

## Competing financial interests

The authors declare no competing financial interests.

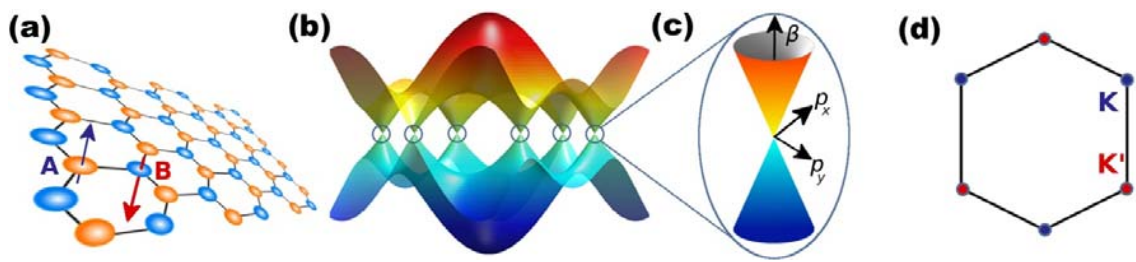

Fig. 1: **Schematic of the lattice and band gap structure of the honeycomb lattice.**
(a) Honeycomb lattice structure of graphene, where arrows illustrate the pseudospin representation of honeycomb sublattice (A or B) degree of freedom. (b) The band gap structure of graphene lattice exhibiting six Dirac points. (c) Zoom-in of the linear dispersion close to one of the Dirac points. (d) The first Brillouin zone of the lattice where the location of two inequivalent corners are marked by K and K'.

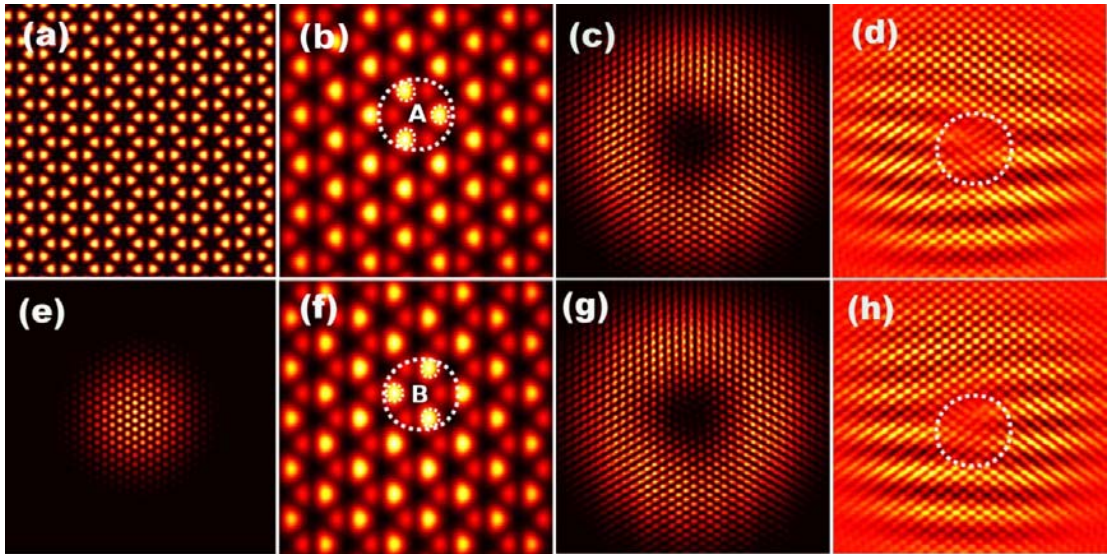

Fig. 2: **Numerical simulation of pseudospin-mediated vortex generation in photonic graphene**. (a) The honeycomb lattice, and (e) the input triangular lattice as a probe formed by three-beam interference. (b, f) Superimposed patterns (zoomed in) when only sublattice A (top row) or B (bottom row) is excited. The output intensity patterns of the probe beam exhibit similar conical diffraction (c, g), but the interferograms obtained with an inclined plane wave reveal opposite phase singularities in the center (d, h) due to excitation of different pseudospin states. (Parameters for simulation are chosen close to those from experiment: the lattice spacing is 7μm, the strength of refractive index modulation is 2×10$^{-4}$, and the propagation distance is 20mm).

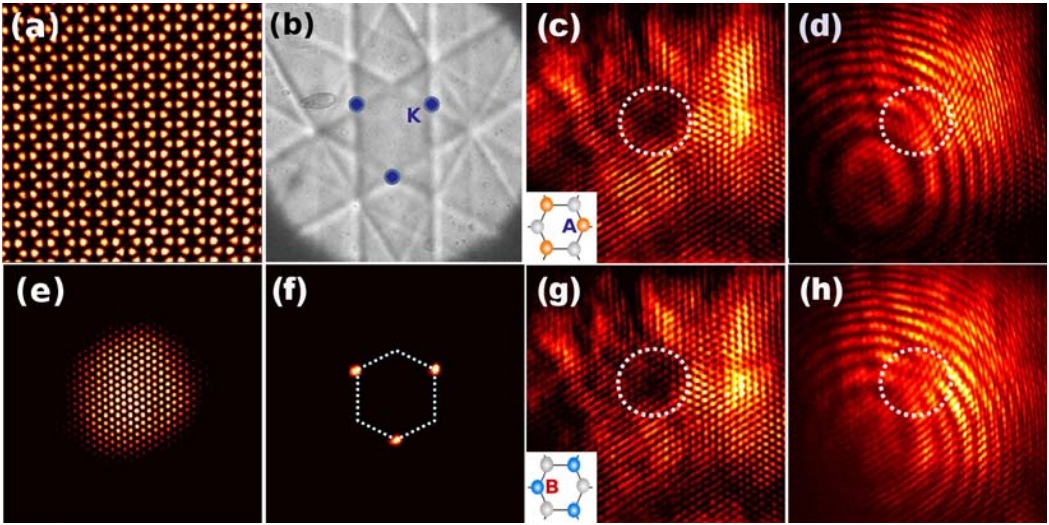

Fig. 3: **Experimental observation of vortex generation by initial excitation of three Dirac points (the K valleys) in photonic graphene corresponding to Fig. 2.** (a) The honeycomb lattice optically induced in a nonlinear crystal, and (e) the probe beam at the input to the lattice containing no phase singularity. (b) Measured Brillouin zone spectrum of the induced lattice, and (f) the *k*-space spectrum of the input beam matching the three marked Dirac points in (b). The white dashed lines mark the first BZ. (c, g) Output intensity patterns when only sublattice A (top row) or B (bottom row) is excited. The inserts illustrate the selective excitation of two sublattices by the probe beam, which leads to opposite vortex singularities as identified from the interferograms (d, h).

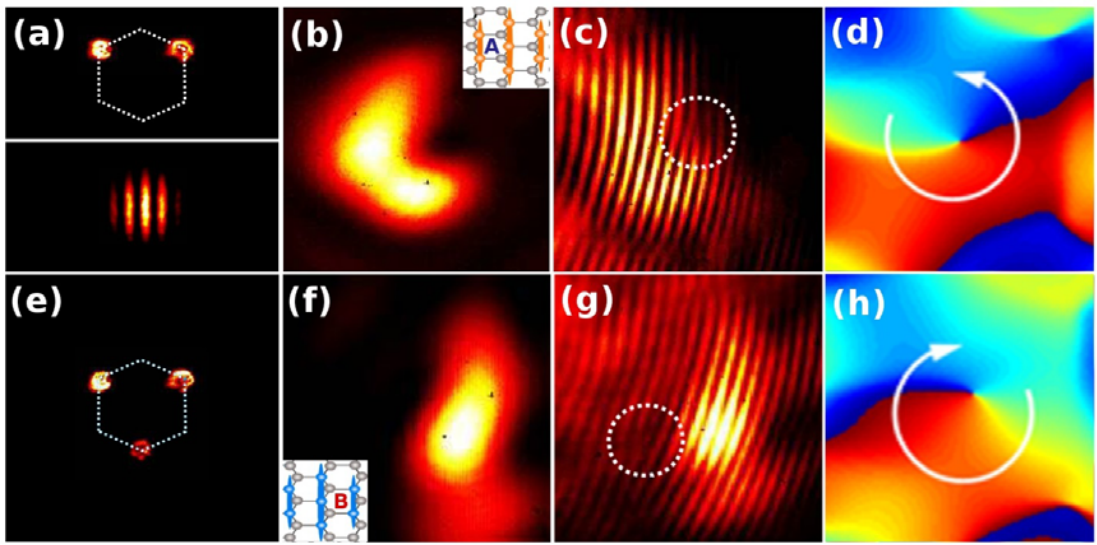

Fig. 4: **Experimental observation of vortex generation by initial excitation of only two Dirac points in photonic graphene.** (a) The input intensity pattern and $k$-space spectrum of the probe beam from two-beam interference, and (e) its output spectrum. Notice the new spectral component appearing at the third Dirac point due to Bragg reflection. (b, f) Output far-field intensity patterns generated at the third Dirac point when only sublattice A (top row) or B (bottom row) is excited. The inserts illustrate the selective excitation of two sublattices by the probe beam. (c, g) Opposite vortex singularities observed from the interferograms of (b, f). (d, h) Output phase structure obtained from corresponding numerical simulation, where the opposite singularities are illustrated by the white arrows.

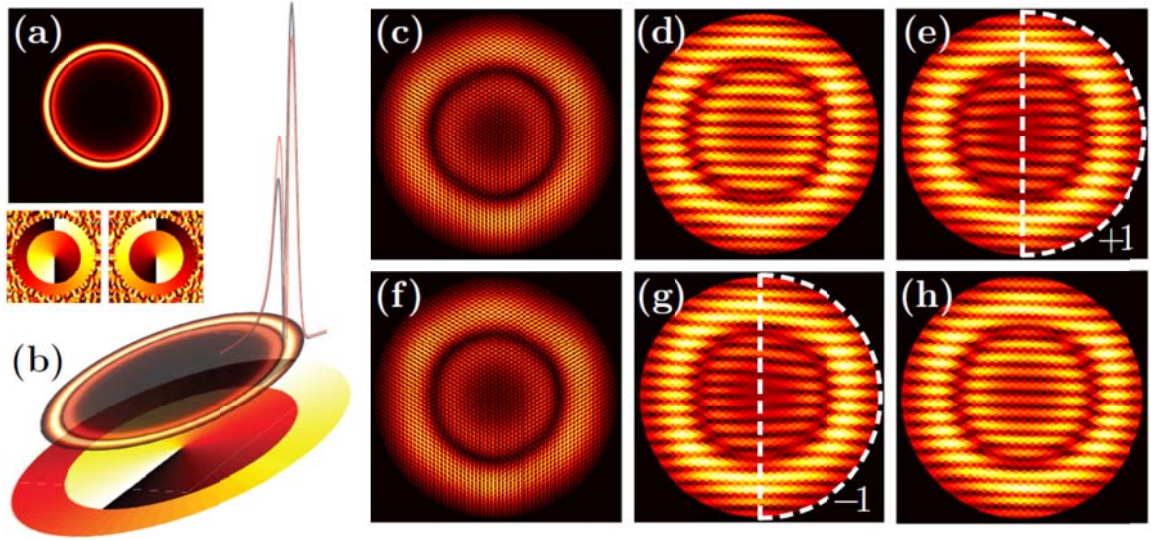

Fig. 5: **Theoretical analysis of the pseudospin-mediated vortex generation.** (a, b) Pseudospin-mediated vortex generation calculated directly from the Dirac equation. When only one of the spinor-components $\psi_A$ ($\psi_B$) is given an initial Gaussian-modulated excitation, the output intensity at $z = 30$ is identical [top panel of (a)] but the corresponding phase is different. The left (right) bottom panel of (a) shows the vortex phase of $\psi_B$ ($\psi_A$) while the other component has uniform phase. Comparison between results from numerical solution (dark) and asymptotic calculation is shown in (b). (c)-(h) Intensity and phase of the output optical field obtained from numerical simulation of the paraxial model with a graphene-type HCL potential based on decomposing the optical field into its spinor components. In the top (bottom) row, only sublattice A (B) is initially excited with a Gaussian modulation, which leads to similar intensity patterns (c, f) but opposite phase structures. The second column (d, g) shows the interferograms of the wavefunction $\Psi_A$ and the third column (e, h) shows the corresponding interferograms of $\Psi_B$. These components $\Psi_A$ and $\Psi_B$ are derived by restricting the paraxial wavefuction $\Psi$ to the discrete locations that are determined by the lattice elements A and B (see the relevant discussion in the text and Supplementary Information section). When the eigenstate of the lattice spin corresponding to A (B) is initially excited, a vortex is generated in the sublattice B (A) with a topological charge +1 (-1), respectively. For illustration purposes we crop the area outside a disk that consists almost solely of the irrelevant plane-wave contribution.